\documentclass[sigconf]{acmart}

\AtBeginDocument{%
  }

\copyrightyear{2024}
\acmYear{2024}
\setcopyright{acmlicensed}\acmConference[SVR 2024]{Symposium on Virtual and Augmented Reality}{September 30-October 3, 2024}{Manaus, Brazil}
\acmBooktitle{Symposium on Virtual and Augmented Reality (SVR 2024), September 30-October 3, 2024, Manaus, Brazil}
\acmDOI{10.1145/3691573.3691618}
\acmISBN{979-8-4007-0979-1/24/09}

\begin{document}

\title{Attention Guidance through Video Script: A Case Study of Object Focusing on 360º VR Video Tours}


\author{Paulo Vitor Santana da Silva}
\affiliation{%
  \institution{AKCIT}
  \institution{Federal University of Goiás}
  \city{Goiânia}
  \state{Goiás}
  \country{Brazil}
}
\email{paulosantana@discente.ufg.br}

\author{Arthur Ricardo Sousa Vitória}
\affiliation{%
  \institution{AKCIT}
  \institution{Federal University of Goiás}
  \city{Goiânia}
  \state{Goiás}
  \country{Brazil}
}
\email{arthurvitoria@discente.ufg.br}

\author{Diogo Fernandes Costa Silva}
\affiliation{%
  \institution{AKCIT}
  \institution{Federal University of Goiás}
  \city{Goiânia}
  \state{Goiás}
  \country{Brazil}
}
\email{diogo_fernandes@egresso.ufg.br}

\author{Arlindo Rodrigues Galvão Filho}
\affiliation{%
  \institution{AKCIT}
  \institution{Federal University of Goiás}
  \city{Goiânia}
  \state{Goiás}
  \country{Brazil}
}
\email{arlindogalvao@ufg.br}

% \affiliation{%

% \affiliation{%

% \affiliation{%

% \affiliation{%

% \affiliation{%

% \affiliation{%

% \affiliation{%

\renewcommand{\shortauthors}{Paulo Vitor S. Silva, Arthur Ricardo S. Vitória, Diogo F. Costa Silva, and Arlindo R. Galvão Filho}

\begin{abstract}
  Within the expansive domain of virtual reality (VR), 360º VR videos immerse viewers in a spherical environment, allowing them to explore and interact with the virtual world from all angles. While this video representation offers unparalleled levels of immersion, it often lacks effective methods to guide viewers' attention toward specific elements within the virtual environment. This paper combines the models Grounding Dino and Segment Anything (SAM) to guide attention by object focusing based on video scripts. As a case study, this work conducts the experiments on a 360º video tour on the University of Reading. The experiment results show that video scripts can improve the user experience in 360º VR Videos Tour by helping in the task of directing the user's attention.
\end{abstract}



\begin{teaserfigure}
\centering
  \includegraphics[width=.8\textwidth]{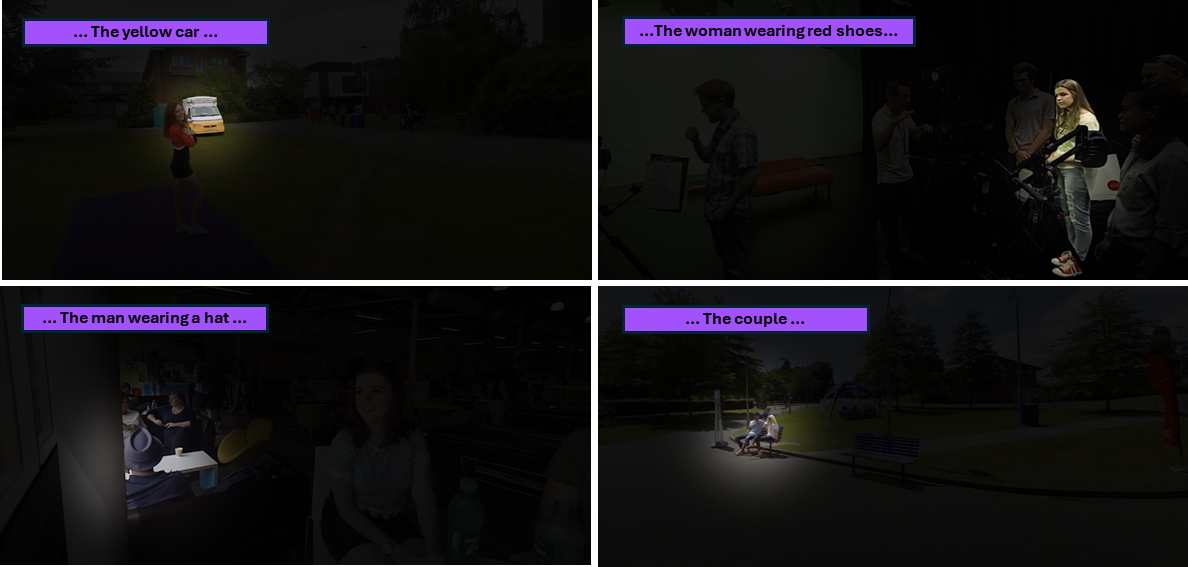}
  \caption{Guiding users' attention on 360º VR video tours through textual scripts using deep learning and computational vision}
  \label{fig:teaser}
\end{teaserfigure}


\maketitle

\begin{figure*}[!htb]
    \centering
    \includegraphics[width=.75\textwidth]{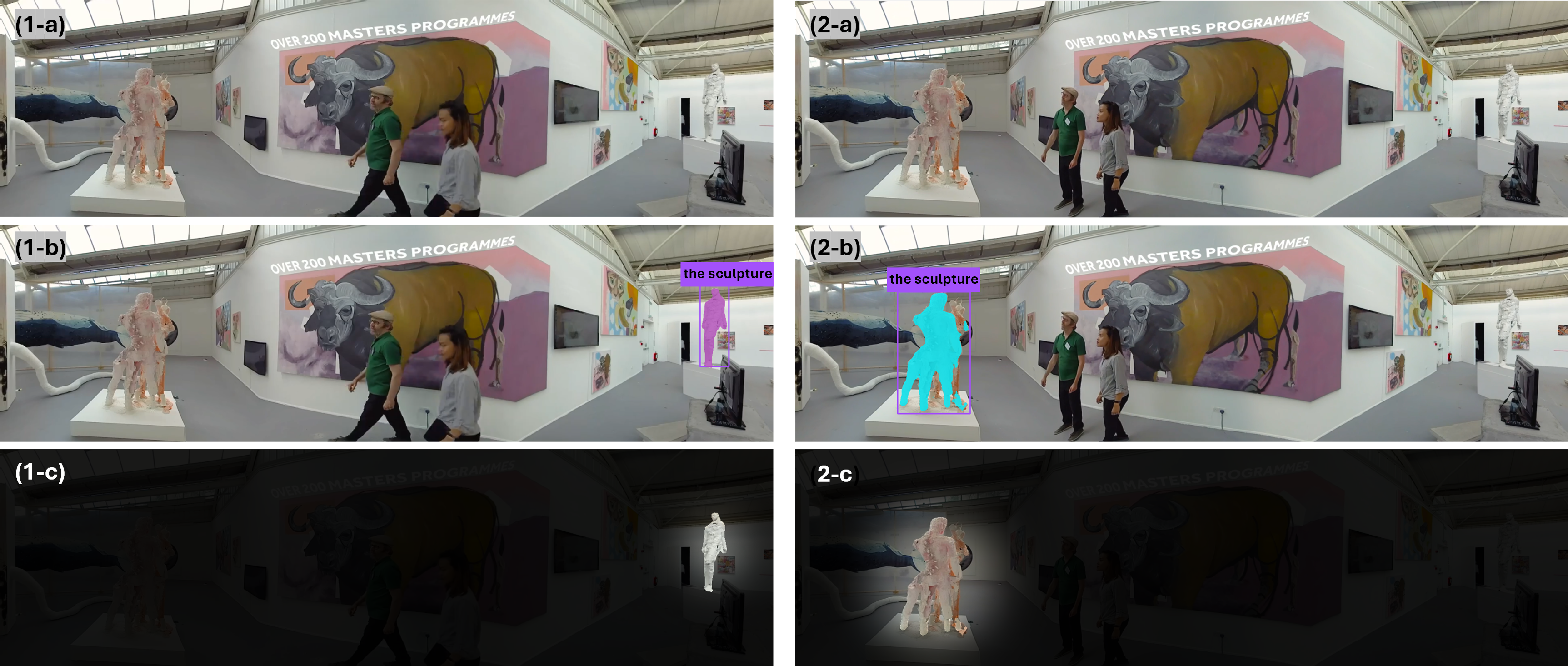}
    \caption{Different moments of the scene showing the museum on the video tour. It is defined in the video script as “Look at the sculpture of a person on the right side” at moment (1) and “Look at the sculpture of a centaur on the left side” at moment (2). (a) The original frame of the video. (b) The object described in the script detected and segmented. (c) The target object with the vignette effect applied.}
    \label{fig:museum-result}
\end{figure*}

\section{Introduction}
Attention guidance involves the interpretation of pertinent information and directing focus towards specific aspects of the visual scene. Due to its immersive viewing experience and omnidirectional view, 360º videos have gained popularity in various areas \cite{liu2017360, macquarrie2017cinematic} such as entertainment, education, and more \cite{guervos2019using, van2017virtual, herault2018using, pulijala2017vr}. Users who are experiencing virtual environments have the freedom to explore the full scene around them, which can lead to a different experience than expected by the narrative \cite{mti6070054}. However, such narratives are still undergoing experiments to establish an effective narrative language \cite{maranes2020exploring}. 

To control the narrative, several techniques can be utilized to captivate the audience's focus on particular elements within a frame \cite{maranes2020exploring}. Depth of field control is a widely used technique in which the focal length of the camera is adjusted to highlight important elements while blurring the rest of the scene \cite{danieau2017attention}. A similar approach can be applied to virtual reality environments \cite{hillaire2007depth}, in which visual blur effects improve the experience of participants. 

% \begin{figure*}[!htb]
%     \centering
%     \includegraphics[width=1\textwidth]{figures/results/museum/museum-grouped-1.png}
%     \caption{Different moments of the scene showing the museum on the video tour. It is defined in the video script as “Look at the sculpture of a person on the right side” at moment (1) and “Look at the sculpture of a centaur on the left side” at moment (2). (a) The original frame of the video. (b) The object described in the script detected and segmented. (c) The target object with the vignette effect applied.}
%     \label{fig:museum-result}
% \end{figure*}

Several works address how to effectively improve how to guide users' attention through visual features \cite{speicher2019exploring, schmitz2020directing}. In the study proposed by Wallgr{\"u}n et al. \cite{wallgrun2020comparison}, they assess the efficacy of three distinct visual guiding mechanisms (arrow, butterfly, and radar) in educational VR tour applications of real-world sites through a user study. All mechanisms showed an improvement over the no-guidance condition, with the arrow having greater preference between 33 participants. 

The work of Danieau et al. \cite{danieau2017attention} proposes an overview of visual resources to guide users through narratives 360º videos, comparing a fade to black, desaturation, blur, and deformation. The intensity of each effect is gradually increased to direct the user's attention toward the area of interest. Moreover, the effectiveness of fade to black and desaturation effects were assessed through a user study, showing promising results yet highlighting the difficulty to make a user unconsciously move his head. In the work of Woodworth et al. \cite{woodworth2023design} 9 distinct visual features are explored in order to guide or restore users' attention. Their work encompasses a guidance task wherein subjects gaze at different regions of interest in a randomized order, and a restoration task in which gaze sequences are interrupted by distraction events, requiring a return of focus. Showing an extensive direct comparison of attention guidance and restoration cues.

Hillaire et al. \cite{4480749} proposed two techniques to enhance user experience during first-person navigation in a virtual environment. These techniques include simulating realistic camera motion similar to human eye movement during walking, and implementing a depth of field blur effect to mimic human perception, where sharp objects are perceived only within a certain range of distances around the focal point. The results showed that participants globally preferred the use of these effects when they are dynamically adapted to the focus point in a virtual environment. 

With the rapid democratization of immersive technologies, numerous design challenges and considerations must be taken into account to improve how the users' experience virtual reality. In this context, this work proposes an approach that, through a video script and deep learning, automatically recognizes important elements in a scene and then guides the user's attention to them through a vignette effect, using VR tours as a study case.  

This work is organized as follows, in Section \ref{sec:mm} the material and methods are described, describing all techniques utilized to generate the regions of interests and segmentation, following by how we apply the visual feature of vignette to guide the user's attention. In Section \ref{sec:results} the experimental results are shown using a 360º video tour on the University of Reading. Finally, in Section \ref{sec:conclusion} the results and implications of the results are discussed on how these mechanisms can enhance attention guidance in virtual reality environments.  

\section{Material and Methods}
\label{sec:mm}
\begin{figure}[!htb]
    \centering
    \includegraphics[width=.6\columnwidth]{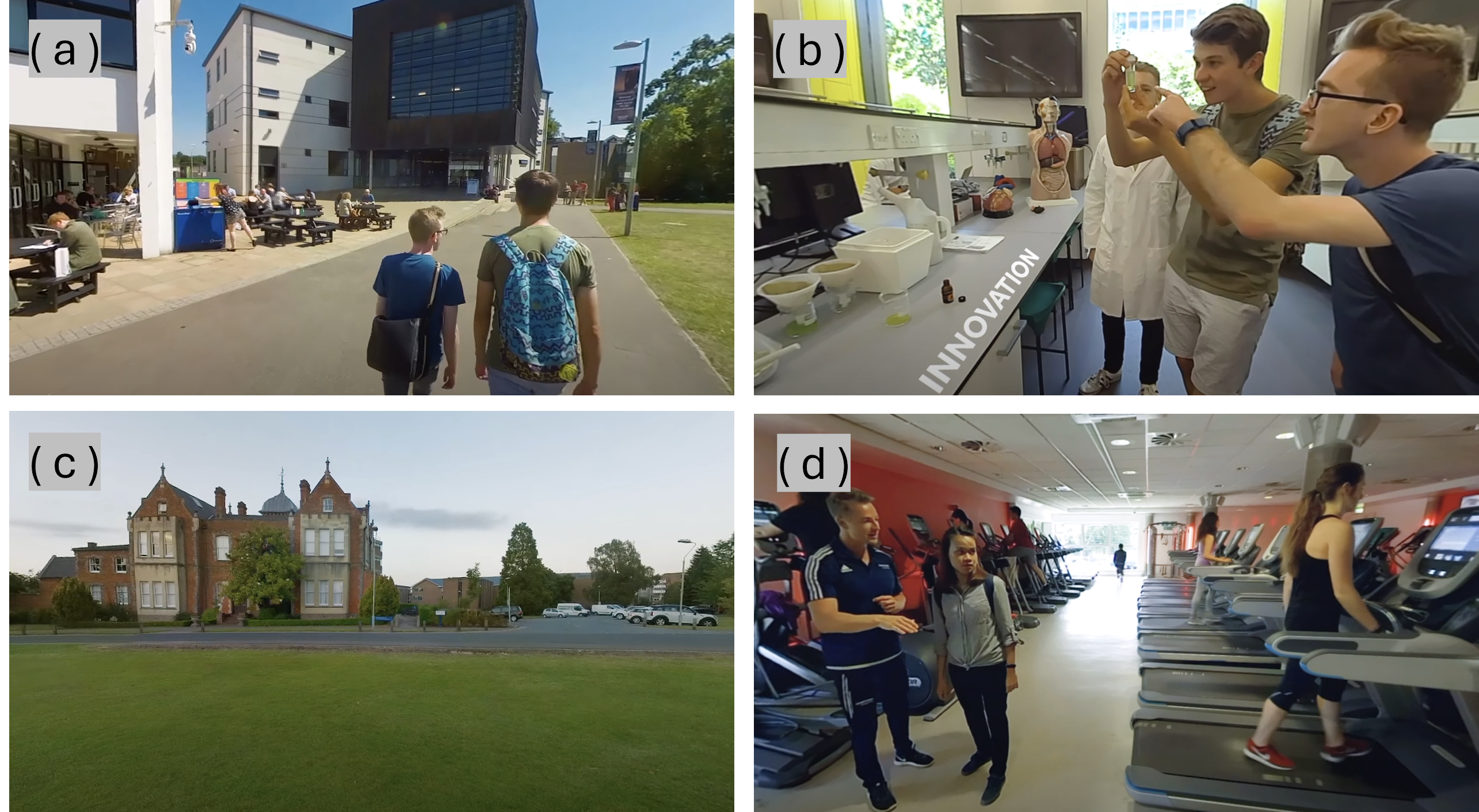}
    \caption{Different frames from the 360º video showcasing distinct environments: (a) depicts an external area, (b) showcases a biology Laboratory, c) shows an external building, and (d) the gym.}
    \label{fig:dataset-examples}
\end{figure}

The 360º videos used consists in a video tour through the campus of University of Reading, featuring both indoor and outdoor environments as shown in Figure \ref{fig:dataset-examples}. During the video, the user is moved to different locations, allowing him to freely look around to several university environments. Moreover, at some frames, there might be more than one region of interest that the user's attention should be directed. 

\begin{figure}[!htb]
    \centering
    \includegraphics[width=.65\columnwidth]{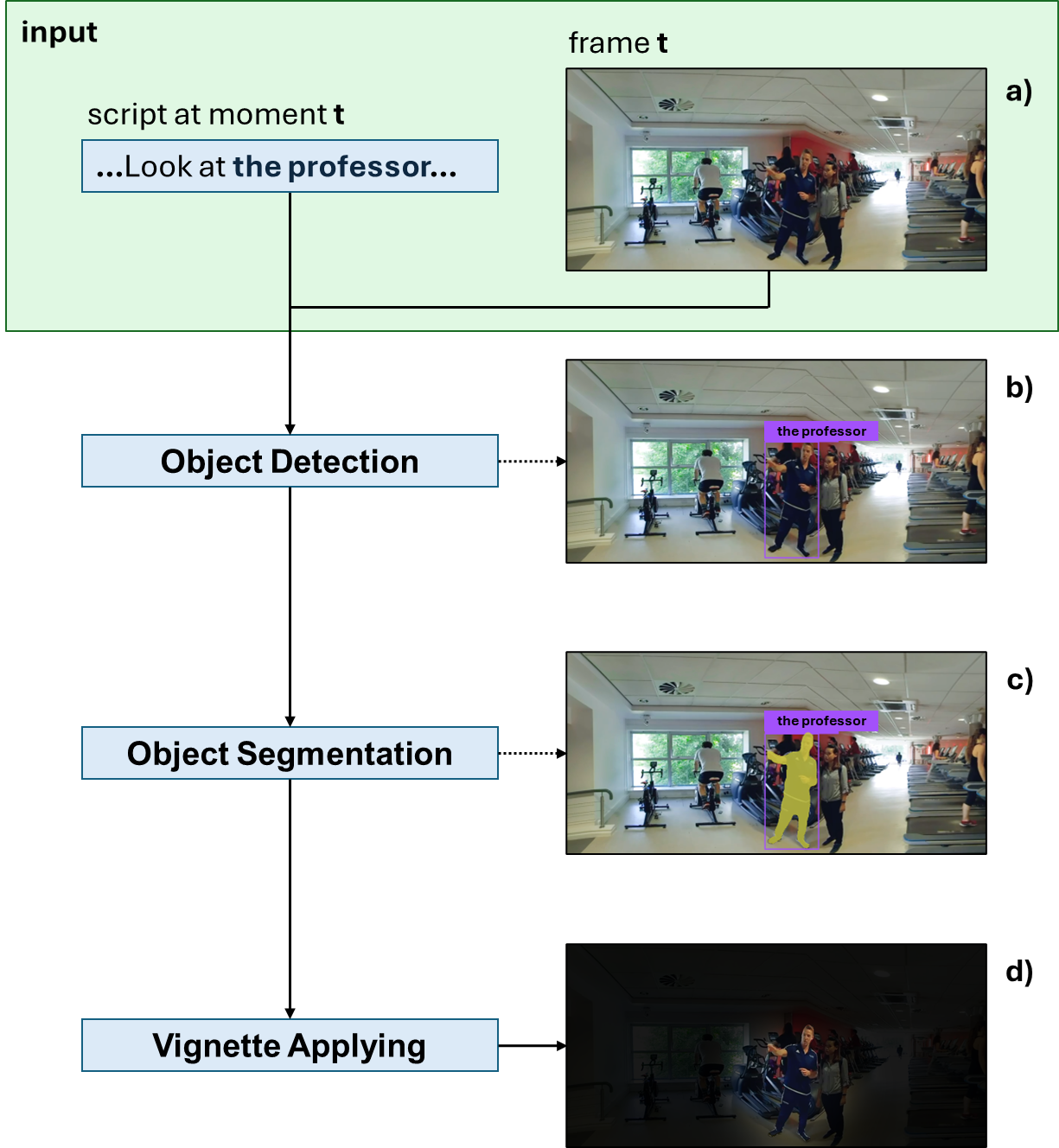}
    \caption{General workflow for a single frame in a 360º VR Tour. a) Through a given input video description along with a selected 360º input frame $t$ as input to Grounding Dino. b) Grounding Dino selects the area (bounding-box) with higher confidence. c) The output bounding-box and image are then used as input to SAM for Object Segmentation, which outputs a segmentation mask. d) Uses the segmentation masks and bounding-box to create a vignette effect that indicates where the user must pay attention.}
    \label{fig:general-workflow}
\end{figure}

Given a video script, containing the target object description in relation to time intervals of the video, on every n frames, the object description and frame are sent as input to Grounding Dino, which in turn, detect the object on the scene and return its bounding-box coordinates as shown in Figure \ref{fig:general-workflow}-b). These coordinates are sent as input jointly with the input frame to the SAM, so that the segmentation mask of the target object can be computed, as shown in Figure \ref{fig:general-workflow}-c). With the bounding-boxes coordinates and the mask of the region of interest, a vignette effect is applied to the scene, indicating where the user must pay attention at this moment as shown in Figure \ref{fig:general-workflow}-d).

\subsection{Object Detection}
\label{sec:gd}

\begin{figure*}[!htb]
    \centering
    \includegraphics[width=.75\textwidth]{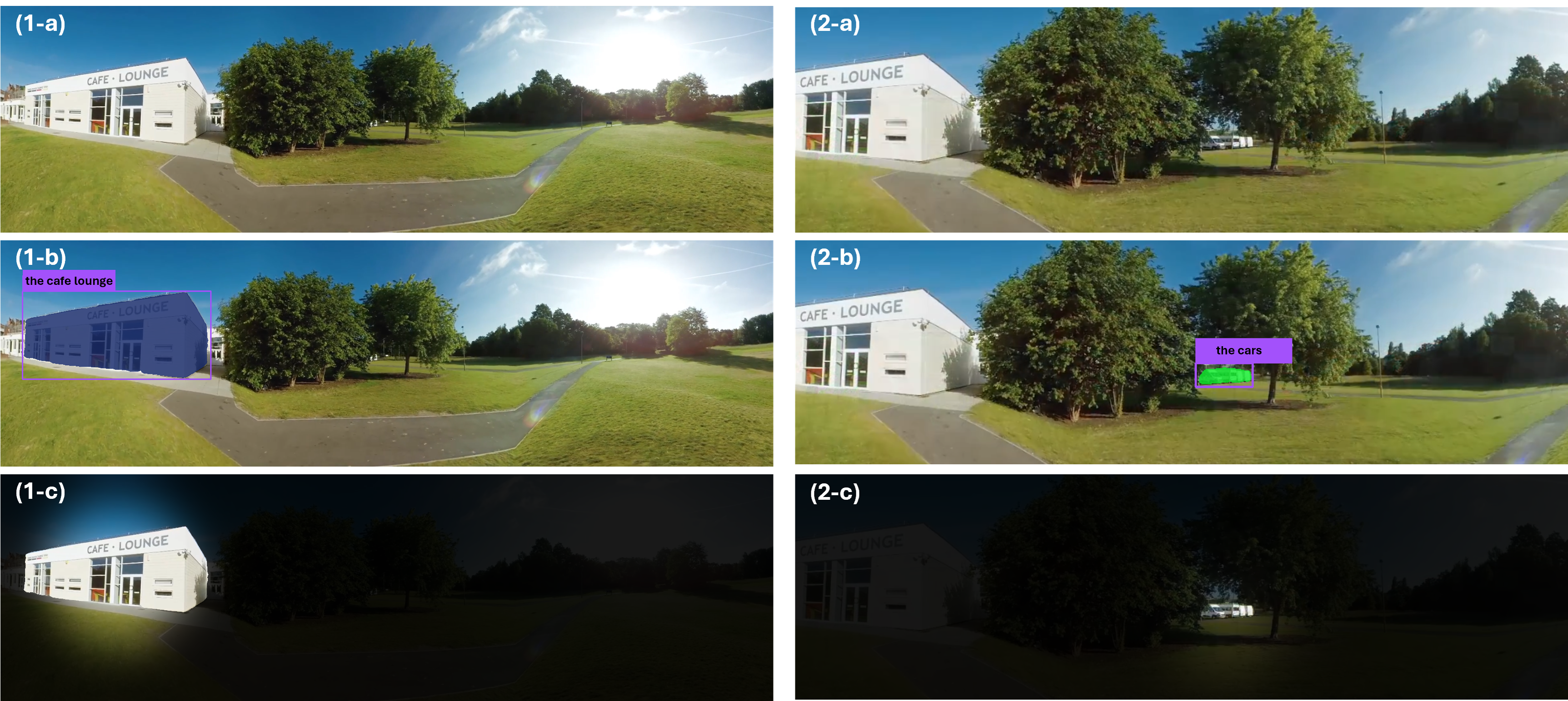}
    \caption{Different moments of the scene showing cafe-lounge on the video tour. It is defined on video script to “Look at the cafe lounge” on moment (1) and “Look at the cars between the trees” on moment (2). (a) The original frame of the video. (b) The object described on the script detected and segmented. (c) The target object with the vignette effect applied.}
    \label{fig:cafe-result}
\end{figure*}

\begin{figure}[!htb]
    \centering
    \includegraphics[width=0.75\columnwidth]{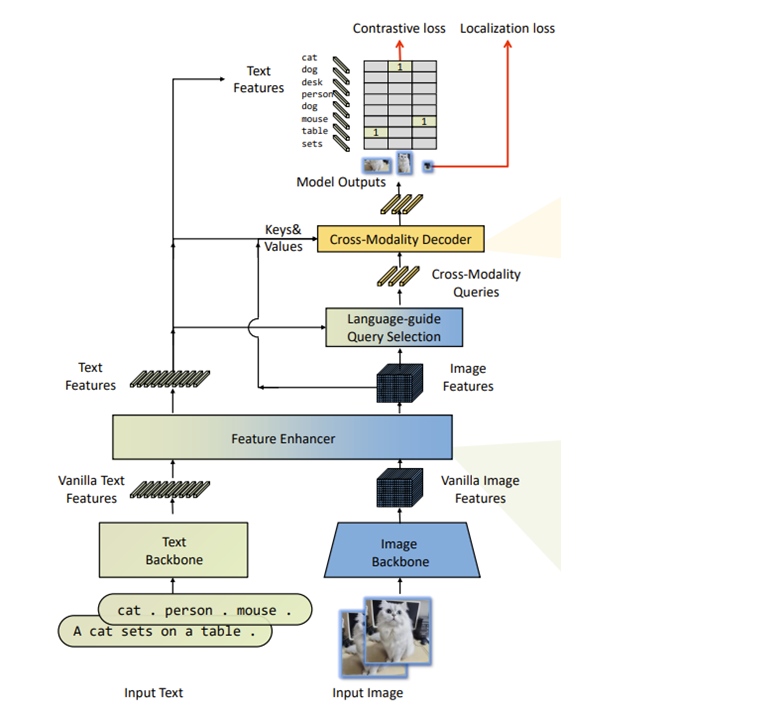}
    \caption{Grounding Dino Architecture  \cite{liu2023grounding} }
    \label{fig:gdino-architecture}
\end{figure}

Grounding Dino \cite{liu2023grounding} is an object detection model based on the Transformer-based detector Dino. Its architecture is composed basically by three components, a feature enhancer, a language-guided query selection and a cross-modality decoder as shown in Figure \ref{fig:gdino-architecture}. For every pair of image and text, the features are extracted using an image backbone and a text backbone, respectively. These two sets of features are then inputted into a feature enhancer module to merge them across modalities. The feature enhancer comprises several layers designed for enhancing features. Deformable self-attention is employed to enrich image features, while standard self-attention is used for text feature enhancement. Once cross-modality text and image features are obtained, a language-guided query selection module is employed to pick cross-modality queries from the image features. These chosen cross-modality queries are then input into a cross-modality decoder. Each cross-modality query is fed into a self-attention layer, an image cross-attention layer to combine image features, a text cross-attention layer to combine text features, and a fully-connected layer in each cross-modality decoder layer. The resulting queries from the final decoder layer are utilized to forecast object boxes and derive corresponding phrases.

The model pre-training strategy involves three key data types, detection data, grounding data and caption data. Inspired by GLIP \cite{glip}, the object detection task is reframed into a phrase grounding task by integrating category names into text prompts sourced from datasets like COCO \cite{coco}, O365 \cite{o365}, and OpenImage (OI) \cite{o365}, with category names dynamically sampled during training for text input variation. For grounding data, GoldG and RefC datasets preprocessed by MDETR \cite{kamath2021mdetr} are utilized, encompassing images from Flickr30k entities \cite{plummer2015flickr30k}, Visual Genome \cite{krishna2017visual}, RefCOCO, RefCOCO+, and RefCOCOg, enabling direct training of Grounding DINO. To enhance model performance, semantically enriched caption data is incorporated using pseudo-labeling, where a proficiently trained model generates these labels, aiding in the understanding of novel categories.

\subsection{Object Segmentation}
\label{sec:sa}

\begin{figure}[!htb]
    \centering
    \includegraphics[width=0.5\columnwidth]{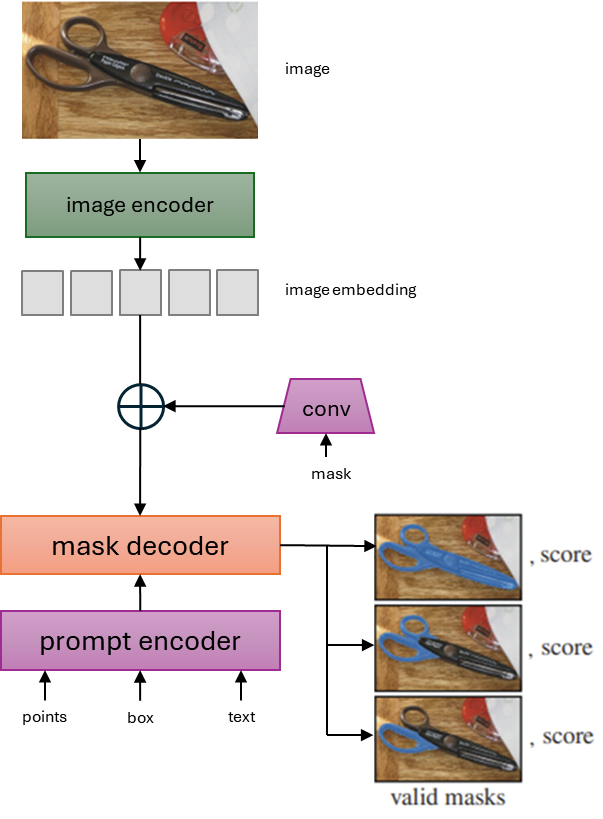}
    \caption{Segment Anything Architecture \cite{Kirillov_2023_ICCV}}
    \label{fig:sam-architecture}
\end{figure}

The Segment Anything Model (SAM) \cite{Kirillov_2023_ICCV} is a versatile segmentation model designed for open-world applications, capable of isolating any object within an image using appropriate prompts such as points, bounding-boxes, or text. Trained on an extensive dataset comprising over 11 million images and 1.1 billion masks, SAM exhibits robust performance even in zero-shot scenarios. However, while its capabilities are significant, the model relies on point or box prompts for accurate object identification, as arbitrary text inputs may not be enough for effective segmentation. Its architecture is composed by three components, an image encoder, a prompt encoder and a mask decoder as shown in Figure \ref{fig:sam-architecture}. The core of the image encoder is a masked auto-encoder, which leverages a vision transformer for scalability. The authors utilized a ViT-H/16, a large-scale vision transformer model designed to handle a 16×16 patch size, featuring a 14×14 windowed attention and four global attention blocks equally spaced apart. The resulting output from this encoder is a feature embedding, downscaled to a 16x smaller version of the original image. This downsizing step is important for efficient processing while preserving essential image characteristics. The model operates on input images with a resolution of 1024×1024×3, and transforms them  into a dense embedding sized 64×64×256. The prompt encoder incorporates two types of prompts: sparse (points, boxes, and text) and dense (masks). Sparse prompts guide the model through representations of points and boxes via learned embeddings, while text prompts utilize CLIP (Contrastive Language-Image Pre-Training) \cite{radford2021learning} without modifications. Dense prompts, represented by masks, maintain spatial correspondence with images. These masks undergo downsizing by a factor of 4 before input, followed by additional downsizing within the model. Gelu activation and layer normalization enhance each layer, and the resulting mask embedding is added to the image embedding. When no mask prompt is provided, a learned embedding for 'no mask' is applied to each image embedding location. The mask decoder is influenced by transformer segmentation models and undergoes modifications to align with transformer decoder framework. The adaptation involves incorporating a learned output token embedding into the prompt embedding prior to decoder processing. This token embedding contains useful information for effective image segmentation, similar to the function of class tokens in vision transformers for image classification. Within each decoder layer, four primary operations occur: self-attention on the tokens, cross-attention from tokens to the image embedding, updates of each token via a point-wise multi-layer perceptron, and cross-attention from the image embedding to tokens. This latter step facilitates the integration of prompt information into the image embedding. During cross-attention, the image embedding is treated as a set of 256-dimensional vectors, enhancing the model's segmentation capabilities.

\section{Results}
\label{sec:results}

Object detection using Grounding Dino was made using zero-shot with a box-threshold equals 0.3 and text-threshold equals 0.25, as recommended by the authors. Although the model may return different bounding boxes, only the one with the highest confidence is considered. It was used the checkpoint ViT-H of the SAM model for object segmentation and the zero shot approach was followed.

On Figures 2 and 5 are shown two moments (1 and 2) from scenes at Reading University, presenting the museum and the cafe-lounge, respectively. According to the video script, at moment (1) in Figure 5, attention should be on “The cafe-lounge,” and at moment (2), on “The cars between the trees.” Similarly, for Figure 2, attention should be on “The sculpture of a person on the right side” at moment (1), and on “The sculpture of a centaur on the left side” at moment (2). Object detection and segmentation were successfully performed for both scenes, as shown in Figures 5 (1-b), 5 (2-b), and 2 (1-b). The vignette effect effectively directed attention to the respective target elements, as seen in Figures 5 (1-c), 5 (2-c), and 2 (1-c). However, in Figure 2 (2-b), SAM failed to correctly segment the sculpture of the centaur, resulting in the vignette effect partially obscuring the target element.

% \textcolor{cyan}{e como contornar esse erro? se for um erro comum comentar nas conclusões como uma limitacao desse método.}.

The observed error occurred due to the utilization of a zero-shot approach during the inference on SAM. Fine-tuning the model on a dataset containing images related to the elements presented in this case study would result in greater effectiveness during the segmentation phase.

\section{Conclusion}
\label{sec:conclusion}

To direct users' attention to specific regions of interest outlined by an input script during a 360º video tour, this study proposes a method that relies on combining the models Grounding Dino for object detection and SAM for object segmentation in scenes. Additionally, the application of a vignette feature is employed to direct users to where their focus is required according to the video script. The experimental results demonstrate that the integration of deep learning methodologies for comprehending video scripts, coupled with computational vision techniques, yields significant enhancements in user experience within 360º VR video tours. In this context, future works will be focused on the improvement of the object segmenting technique by fine-tuning SAM in a dataset containing images related to elements presented in this case study. Besides, a study will be conducted through interviews about the effectiveness of the proposed method in directing users' attention.

\begin{acks}
This work has been fully/partially funded by the project Research and Development of Algorithms for Construction of Digital Human Technological Components supported by Advanced Knowledge Center in Immersive Technologies (AKCIT), with financial resources from the PPI IoT/Manufatura 4.0 / PPI HardwareBR of the MCTI grant number 057/2023, signed with EMBRAPII
\end{acks}

% Despite the limitations of the object detector and object segmenting model, this method can be a good alternative when it comes to improve users' attention during a 360º VR video tour. 
\bibliographystyle{ACM-Reference-Format}
\bibliography{references}

\end{document}